\documentclass[sigplan, nonacm]{acmart}

\setcopyright{none}
\acmDOI{}
\acmISBN{}
\acmConference[Preprint]{Preprint}{}{2026}{}
\copyrightyear{}
\acmYear{}
\usepackage{array}
\usepackage{amsmath,amsfonts}
\usepackage{algorithmic}
\usepackage{graphicx}
\usepackage{textcomp}
\usepackage{xcolor}
\usepackage{minted}
\usepackage{listings}
\usepackage[utf8]{inputenc}
\usepackage{enumitem}
\usepackage{multirow}
\usepackage{listings}
\usepackage{wrapfig}
\usepackage{adjustbox}

\newcommand{\pname}{\texttt{ApproxHDC}}

\definecolor{keyword}{RGB}{127, 0, 170}    % purple for types
\definecolor{comment}{RGB}{106, 115, 125}  % gray for comments
\definecolor{function}{RGB}{0, 55, 120}    % dark blue for functions
\definecolor{number}{RGB}{0, 128, 0}       % green for numbers
\definecolor{bg}{RGB}{246, 248, 250}       % light gray background

\lstdefinelanguage{HDCpp}{
  morekeywords={hypervector, hypermatrix, int, void},
  morekeywords=[2]{__hetero_hdc_matmul, __hetero_hdc_hamming_distance,
    __hetero_hdc_red_perf, __hetero_hdc_arg_min},
  morecomment=[l]{//},
  morestring=[b]",
}

\definecolor{codegreen}{rgb}{0,0.6,0}
\definecolor{codegray}{rgb}{0.5,0.5,0.5}
\definecolor{codepurple}{rgb}{0.64,0.08,0.08}
\definecolor{backcolour}{rgb}{0.95,0.95,0.92}

\lstdefinestyle{mystyle}{
    backgroundcolor=\color{bg},   
    commentstyle=\color{codegreen},
    keywordstyle=\color{magenta},
    numberstyle=\tiny\color{codegray},
    stringstyle=\color{codepurple},
    basicstyle=\ttfamily\footnotesize,
    breakatwhitespace=false,         
    breaklines=true,                 
    captionpos=b,                    
    keepspaces=true,                 
    numbers=left,                    
    numbersep=5pt,                  
    showspaces=false,                
    showstringspaces=false,
    showtabs=false,                  
    tabsize=2
}

\begin{document}

\title{Compiler-Driven Approximation Tuning for Hyperdimensional Computing}

\author{Xavier Routh}
\affiliation{%
  \institution{University of Illinois Urbana-Champaign}
  \city{Urbana}
  \state{IL}
  \country{USA}
}
\email{xrouth2@illinois.edu}

\author{Abdul Rafae Noor}
\affiliation{%
  \institution{University of Illinois Urbana-Champaign}
  \city{Urbana}
  \state{IL}
  \country{USA}
}
\email{arnoor2@illinois.edu}

\author{Akash Kothari}
\affiliation{%
  \institution{University of Illinois Urbana-Champaign}
  \city{Urbana}
  \state{IL}
  \country{USA}
}
\email{akashk4@illinois.edu}

\author{Zheyu Li}
\affiliation{%
  \institution{University of California San Diego}
  \city{La Jolla}
  \state{CA}
  \country{USA}
}
\email{zhl178@ucsd.edu}

\author{Mahbod Afarin}
\affiliation{%
  \institution{University of California San Diego}
  \city{La Jolla}
  \state{CA}
  \country{USA}
}
\email{mafarin@ucsd.edu}

\author{Tajana Rosing}
\affiliation{%
  \institution{University of California San Diego}
  \city{La Jolla}
  \state{CA}
  \country{USA}
}
\email{tajana@ucsd.edu}

\author{Vikram Adve}
\affiliation{%
  \institution{University of Illinois Urbana-Champaign}
  \city{Urbana}
  \state{IL}
  \country{USA}
}
\email{vadve@illinois.edu}

%%%%%% -- PAPER CONTENT STARTS-- %%%%%%%%

\begin{abstract}

As Moore’s law reaches its physical and economic limits, domain-specific approaches are increasingly employed to accelerate machine learning workloads. Hyperdimensional Computing (HDC) represents one such emerging paradigm, offering an alternative to conventional deep learning techniques. Rooted in cognitive models of computation, HDC is designed bottom-up with hardware efficiency as a first-class objective. Consequently, HDC workloads map naturally to heterogeneous hardware platforms, including CPUs, GPUs, and FPGAs, as well as emerging in-memory computing technologies such as Resistive RAM (ReRAM) and Phase-Change Memory (PCM). The intrinsic tolerance of HDC algorithms to noise and approximation enables substantial performance gains with minimal loss in inference quality. In this work, we introduce \textit{ApproxHDC}, a framework for automated identification and application of domain-specific approximations in HDC workloads. ApproxHDC extends the HPVM-HDC compiler infrastructure to enable retargetable compilation across diverse hardware backends, including CPUs, GPUs, and simulated ReRAM and PCM-based accelerators. The space of possible software and hardware approximations in HDC is exponentially large, making exhaustive exploration intractable; ApproxHDC employs efficient search and analysis techniques to navigate this space and identify high-impact approximations. We further integrate hardware-level approximation mechanisms within the HDC accelerator to demonstrate ApproxHDC’s ability to exploit both software and hardware approximations. Experimental evaluation shows that ApproxHDC achieves up to $\textbf{17.25}\times$ speedup on CPUs, $\textbf{15.02}\times$ on GPUs, and $\textbf{4.69}\times$ on SpecPCM while staying within user-specified quality-of-service constraints, outperforming the state-of-the-art MicroHD framework by $\textbf{3.49}\times$. HDC-specific search space pruning reduces the design space by up to \textbf{86} orders of magnitude, enabling discovery of high-quality configurations within minutes.

\end{abstract}
%\begin{CCSXML}
%<ccs2012>
%  <concept_id>00000000.0000000.0000000</concept_id>
%  <concept_desc>Do Not Use This Code, Generate the Correct Terms for Your Paper</concept_desc>
%  <concept_significance>500</concept_significance>
%  %<concept_id>00000000.00000000.00000000</concept_id>
%  <concept_desc>Do Not Use This Code, Generate the Correct Terms for Your Paper</concept_desc>
%  <concept_significance>300</concept_significance>
%  %<concept_id>00000000.00000000.00000000</concept_id>
%  <concept_desc>Do Not Use This Code, Generate the Correct Terms for Your Paper</concept_desc>
%  <concept_significance>100</concept_significance>
 % <concept_id>00000000.00000000.00000000</concept_id>
%  <concept_desc>Do Not Use This Code, Generate the Correct Terms for Your Paper</concept_desc>
%  <concept_significance>100</concept_significance>
%</ccs2012>
%\end{CCSXML}

%\ccsdesc[500]{Do Not Use This Code~Generate the Correct Terms for Your Paper}
%\ccsdesc[300]{Do Not Use This Code~Generate the Correct Terms for Your Paper}
%\ccsdesc{Do Not Use This Code~Generate the Correct Terms for Your Paper}
%\ccsdesc[100]{Do Not Use This Code~Generate the Correct Terms for Your Paper}

\keywords{Compilers, Hyperdimensional Computing, Heterogeneous Systems, Approximate Computing}

\maketitle

\section{Introduction}
\label{sec:intro}

With machine learning models pressing up against the memory and computational limits of current hardware, modern hardware devices are experiencing reduced performance and higher energy consumption \cite{hpvm-hdc}. To address the performance, energy, and accuracy needs of modern workloads, the brain-inspired paradigm \textit{Hyperdimensional Computing (HDC)} \cite{kanerva2009hyperdimensional} is attracting increasing interest. HDC encodes information as \textit{hypervectors}, very high-dimensional vectors made up of thousands of elements. Because HDC operations are lightweight, highly parallelizable, and noise-tolerant, they are particularly well-suited for computations in resource-constrained environments. Consequently, HDC has been employed in a wide range of applications, including natural language processing \cite{rahimi2016robust}, robotics \cite{mitrokhin2019learning}, voice and gesture recognition \cite{imani2017voicehd, rahimi2016hyperdimensional}, emotion recognition \cite{chang2019hyperdimensional}, graph and hypergraph learning \cite{nunes2022graphhd, HygHD}, DNA sequencing \cite{kim2020geniehd}, recommended systems \cite{guo2021hyperrec}, and bio-signal processing \cite{asgarinejad2020detection}.

A key property of HDC is its intrinsic resilience to noise and approximation. Because information is distributed across thousands of dimensions, individual errors in computation or storage barely affect the final result \cite{SurveyHDC}. This stands in sharp contrast to deep neural networks, where even small perturbations in intermediate activations can cascade into large accuracy losses. The error tolerance of HDC opens a rich design space, allowing one to deliberately introduce imprecision through reduced arithmetic precision, skipped loop iterations, cheaper algorithmic variants, or relaxed hardware operating points while still maintaining acceptable inference quality. These approximations, in principle, translate directly into gains in execution time and energy consumption.

The challenge, however, is that the space of possible approximations is combinatorially large. A single HDC application may contain dozens of primitive operations, each offering multiple approximation strategies with continuous or discrete parameters. When the application targets heterogeneous hardware, the space expands further, because each backend exposes its own set of hardware-level knobs such as ADC resolution, write-verify depth, multi-level cell configuration, and analog quantization scale. Manually navigating this space is impractical, as configurations rarely transfer across datasets or backends, and cross-operation interactions are hard to predict without empirical evaluation.

Existing approximation frameworks partially address this problem but fall short in the HDC context. General-purpose approximate computing systems such as EnerJ \cite{enerJ}, ApproxHPVM \cite{approxhpvm}, and ApproxTuner \cite{approxtuner} target deep learning and image processing pipelines and lack the domain-specific compiler support needed to analyze HDC primitives. On the HDC side, MicroHD \cite{microhd} explores a limited hyperparameter space using a Python library without compiler integration, making it unable to exploit hardware-specific approximation knobs or to apply fine-grained, per-operation approximations. Neither class of prior work jointly considers software-level and hardware-level approximation for HDC across heterogeneous targets.

In this work, we present \pname{}, a framework that closes this gap. \pname{} extends the \textit{HPVM-HDC} compiler infrastructure \cite{hpvm-hdc} to automatically identify all HDC primitive invocations within an application, construct the applicable approximation design space per operation, and efficiently search that space subject to a user-specified \textit{quality-of-service (QoS)} constraint. The framework supports retargetable compilation to CPUs, GPUs, and simulated ReRAM- and PCM-based in-memory accelerators~\cite{ReRAMAcc, SpecPCM2024}, and it treats both software transformations  and hardware-level knobs  as first-class dimensions of the search space. To keep search tractable, \pname{} provides mechanisms for developers to prune the space through lightweight source annotations, and it leverages domain knowledge about HDC to eliminate unpromising regions productively.

\vspace{0.05in}
\noindent
Overall, this paper makes the following contributions:

\begin{itemize}
\item  We introduce \pname{}, an auto-tuning system for HDC applications that automates approximation selection to jointly optimize end-to-end performance, energy, and quality of service across heterogeneous hardware targets.
\item  We provide unified support for software and hardware approximations in HDC, including integration of in-memory accelerator knobs into the compiler-driven tuning loop.
\item We develop HDC-specific techniques to prune the approximation search space, combining automatic domain-aware pruning with developer-guided annotations, to substantially reduce auto-tuning time. Our pruning reduces the search space by up to \textbf{86}$\times$, enabling discovery of high-quality configurations within minutes. Notably, no single pruning strategy uniformly achieves the highest performance across all applications, underscoring the importance of providing developers with mechanisms to productively guide the search.
\item We evaluate \pname{} extensively across four HDC benchmarks (HD-Classification, HD-Clustering, RelHD, and HD-Hashtable), multiple datasets, and a range of hardware targets. Our results demonstrate: (i) up to $\textbf{17.25}\times$ speedup on CPUs, $\textbf{15.02}\times$ on GPUs, and $\textbf{4.69}\times$ on SpecPCM while staying within user-specified QoS constraints; and (ii) a $\textbf{3.49}\times$ speedup advantage over the state-of-the-art MicroHD framework through a richer, compiler-driven approximation spaces.

% (iii) consistent tuning behavior across different datasets for the same application; (iv) analysis of how approximation configurations transfer across backends (GPU, CPU, PCM), quantifying the need for re-tuning.}
% %\vspace{-1em}

% and (v) a thorough characterization of the performance and energy trade-offs exposed by hardware approximation knobs in PCM and ReRAM devices. 

% \rafae{Once we have all the results, we should also elaborate that no one pruning strategy is uniformly achieving the highest performance; different applications converge differently hence why being able to productively guide the pruning is important.}

\end{itemize}

% \begin{itemize}
% \item An auto-tuning system for HDC applications, automating approximation selection to optimize for end to end application metrics such as (across) performance, energy, and QoS.
% \item Support for both hardware and software approximations in HDC applications.
% \item Experimental evaluation of the tuning system for 5 benchmarks (HD-Classification, HD-Clustering, HyperOMS, RelHD, and HD-Hashtable) across multiple datasets and hardware targets (when applicable). Showing:
% \begin{itemize}
% 	\item Performance, energy and QoS, tradeoffs for whole-application tuning.
% 	\item Evaluation of tuning flexibility / consistency across datasets (for same application).
% 	\item Evaluation of tuning flexibility / consistency across backends (for same application).
% 		- I.e how do you have to re-tune when moving from GPU, CPU, to PCM
% 	\item A thorough evaluation of hardware approximation knobs exposed by PCM and ReRam devices for HDC applications. (Perf and energy numbers)
% 	\item Significant performance gains with marginal loss in QoS.
% \end{itemize}
% \item HDC specific techniques to prune the search space to reduce auto-tuning time. Both automatic and developer guided.
% \end{itemize}
\vspace{-1em}
\section{Background}

\subsection{Hyperdimensional Computing}

Hyperdimensional Computing (HDC) is an emerging computational paradigm inspired by the brain’s use of distributed, high-dimensional representations for cognition. Instead of operating on low-dimensional numerical values as in conventional machine learning, HDC represents data using high-dimensional vectors, called hypervectors, often consisting of thousands of dimensions. Information is encoded and manipulated through simple, mathematically well-defined operations on these hypervectors, enabling highly efficient and robust computation. HDC systems are naturally error-tolerant due to their distributed structure and can learn from limited data without the need for expensive, iterative optimization methods like backpropagation used in deep learning. This makes HDC particularly suitable for resource-constrained environments and applications requiring fast, on-device learning or inference. While traditional deep learning models often require large amounts of data, substantial compute resources, and energy-intensive training pipelines, HDC offers a lightweight but effective alternative with lower computational overhead, faster training, and inherent suitability for hardware acceleration. These properties make HDC a compelling approach for applications in edge AI, biomedical signal processing, and other domains where efficiency, robustness, and interpretability are critical. 

The increased performance and inherent error resilience of HDC come at a cost: achieving the desired accuracy for a given application typically requires extensive exploration of hyperparameters and algorithmic strategies to compensate for the simplified learning mechanisms employed by HDC.
A typical HDC application consists of three main stages: encoding, training, and inference. In the encoding stage, raw input data is transformed into high-dimensional hypervectors, capturing the essential features of the data in a distributed representation. During training, hypervectors corresponding to inputs from the same class are combined to form class hypervectors or prototype representations, typically using simple aggregation operations such as vector addition and normalization. Finally, in the inference stage, a query hypervector is compared to the stored class hypervectors using a similarity or dissimilarity metric, commonly cosine similarity or Hamming distance, to identify the closest match and produce the predicted label. Each of these stages is highly parallelizable, allowing HDC applications to efficiently leverage modern heterogeneous hardware, and the distributed nature of hypervectors provides robustness to noise and approximation throughout the pipeline.

\subsection{HDC++}
\label{bg:hdc++}

HDC++~\cite{hpvm-hdc} is a domain-specific, high-level programming language for hyperdimensional computing, embedded within C++. It exposes core HDC operations as first-class primitives, enabling productive and expressive development of HDC applications. Computations in HDC++ operate on hypervectors, representing individual encoded features or class data, and on hypermatrices, representing collections of hypervectors. Unlike prior programming approaches for HDC, HDC++ supports retargetable compilation across a wide range of hardware, including CPUs, GPUs (with optimized libraries), and custom HDC accelerators, using the HPVM-HDC compiler, a retargetable compiler developed specifically for heterogeneous HDC workloads. In terms of development effort, HDC++ applications typically require about twice the lines of code compared to an equivalent Python implementation targeting CPUs, but are significantly more concise than equivalent CUDA implementations. Crucially, a single HDC++ application can be compiled for multiple hardware targets, including accelerators for which no prior implementation exists, eliminating the need for maintaining multiple target-specific versions of the same application.

Another key feature of HDC++ is the ability for developers to control software approximation through concise source code annotations, typically requiring only one or two lines per HDC primitive. These annotations instruct the HPVM-HDC compiler to generate hardware-specific code with the specified approximation transformations applied. The details of these transformations are described in Section~\ref{sec:hdc-approximations}. This compact annotation mechanism enables productive exploration of the approximation space inherent in HDC, allowing developers to achieve the desired accuracy manually while maintaining control over performance and efficiency trade-offs. Listing~\ref{lst:hdcpp} describes a simple HDC++ example which encodes input data using random projection encoding, calculates the dissimilarity using Hamming distance and then identifies the most dissimilar label using argument minimum. The approximation annotation on line 12 instructs the HPVM-HDC compiler to generate target-specific code which calculates Hamming distance using only the first 1024 elements of data.

% \lstinputlisting[label={lst:hdcpp}, caption={HDC++ example for random-projection encoding and dissimilarity between query and class hypervectors.}]{code_snippets/hdcpp_snippet.tex}

\begin{lstlisting}[label={lst:hdcpp}, caption={HDC++ example for random-projection encoding and dissimilarity between query and class hypervectors.}]
hypervector<617> input_features = ...;
hypermatrix<2048, 617> rp_matrix = ...;
hypermatrix<26, 2048>  clusters = ...;
// Encode via random projection
hypervector<2048> encoded_data =
    __hetero_hdc_matmul(input_features, rp_matrix);
// Compute Hamming distance to all clusters
hypervector<26> dists =
    __hetero_hdc_hamming_distance(encoded_data, clusters);
// Approximate: use only first 1024 dimensions
__hetero_hdc_red_perf(dists, 0.0, 0.5, 1);
// Select nearest cluster
int label = __hetero_hdc_arg_min(dists);
\end{lstlisting}

\subsection{HPVM-HDC}
\label{bg:hpvm-hdc}
HPVM-HDC~\cite{hpvm-hdc} is a retargetable compiler infrastructure designed specifically for hyperdimensional computing (HDC) applications. Built on top of the HPVM~\cite{kotsifakou2018hpvm, ejjeh2022hpvm} heterogeneous compiler framework, HPVM-HDC provides an intermediate representation (IR) that captures the high-level semantics of HDC operations, including encoding, training, and similarity-based inference. The HPVM-HDC IR abstracts away hardware-specific details while preserving the structure of HDC primitives, enabling the compiler to perform optimizations such as parallelization and target-specific code generation. 
Each HDC++ language primitive corresponds to a unique HPVM-HDC IR operation, allowing the intermediate representation to retain the domain-specific semantics of HDC within the compiler. Internally, the HPVM-HDC compiler translates HDC++ applications into the HPVM IR, which provides constructs for representing multiple forms of parallelism, including task-level, data-level, and streaming parallelism. Parallel programs are represented as a hierarchical directed acyclic graph (DAG), where nodes fall into two categories: internal nodes, which capture hierarchical parallelism by containing entire subgraphs, and leaf nodes, which represent individual units of computation expressed in LLVM IR. Edges between nodes denote logical data transfers, meaning explicit copies may or may not be required. Additionally, each node can be annotated with one or more hardware targets, enabling the compiler to generate code for multiple backends and fully exploit heterogeneous platforms such as CPUs, GPUs, and (Intel) FPGAs.

In addition to generating HPVM IR for retargetable heterogeneous compilation, HPVM-HDC implements two parameterized software approximation transformations at the IR level. The first, Automatic Binarization Propagation, performs a quantization-like transformation on hypervectors, representing the resulting data as single-bit packed vectors and inter-procedurally updates computations and allocations as required. The second, Reduction Perforation, applies parameterized loop perforation by regularly skipping iterations in HDC operations that involve reductions. Both of these software approximations are described in detail in Section~\ref{sec:hdc-approximations}. HPVM-HDC also supports compilation for specialized HDC accelerators, including an HD-ASIC and an HD Resistive RAM (ReRAM) accelerator. However, it does not currently exploit hardware-specific approximation strategies for these accelerators, a capability we introduce in this work.

\subsection{ReRAM and SpecPCM}

Resistive RAM (ReRAM) and phase-change memory (PCM) are two classes of non-volatile memory that support in-memory computing, the execution of arithmetic directly within the memory array without moving data to a separate processing unit. Both technologies store information as analog resistance states in dense crossbar arrays, and both can perform vector operations such as dot products and Hamming distance through the physics of current summation along array columns. These properties make them a natural substrate for HDC, where the core computational patterns reduce to parallel element-wise and reduction operations over long vectors.

\textbf{ReRAM-based HDC acceleration.} In a ReRAM crossbar, each memristive device is programmed to a conductance proportional to a hypervector element. A read operation applies input voltages to the rows and senses the summed column currents, effectively computing a dot product in a single analog step. FSL-HD \cite{ReRAMAcc} exploits this structure as a few-shot learning accelerator that performs both tensorized encoding and Hamming-distance inference entirely within the ReRAM array. By keeping hypervectors resident in memory and avoiding off-chip data movement, FSL-HD achieves substantial speedups and energy savings over CPU and GPU baselines, while also improving on the area efficiency of the earlier SAPIENS design \cite{SAPIENS}. The key insight from an approximation standpoint is that ReRAM reads are inherently imprecise: device-to-device variability, write noise, and limited ADC resolution all introduce errors, yet HDC accuracy is largely preserved because the final prediction depends on aggregate statistics across thousands of dimensions rather than on any single element.

\textbf{PCM-based HDC acceleration.} SpecPCM \cite{SpecPCM2024} takes a complementary approach using phase-change memory. The architecture offloads distance computations, spectral clustering distances and Hamming similarity for database search into PCM crossbar arrays organized as parallel memory banks with a 2T2R cell structure. SpecPCM integrates two distinct superlattice PCM device types within the same accelerator: one optimized for fast, low-voltage programming suited to clustering workloads, and another providing long retention and low read noise for reliable similarity search. SpecPCM further increases effective storage density through multi-level cells (MLC), packing multiple hypervector dimensions into a single device using a dimension-packing scheme. Stages not suitable for analog execution (e.g., argument selection) are handled by near-memory ASIC logic, enabling significant speedups and energy-efficiency gains over software baselines.

ReRAM and PCM accelerators provide tunable parameters (e.g., ADC resolution, write-verify depth, MLC levels, quantization scale) that trade computation accuracy for latency, energy, and area by introducing controlled noise. Due to HDC’s robustness to noise, this has minimal impact on accuracy. In this work, \pname{} includes these hardware knobs in its approximation search space alongside software techniques to jointly optimize accuracy, performance, and energy.

\section{Design}
\begin{figure*}
    \centering
    \includegraphics[width=0.75\linewidth]{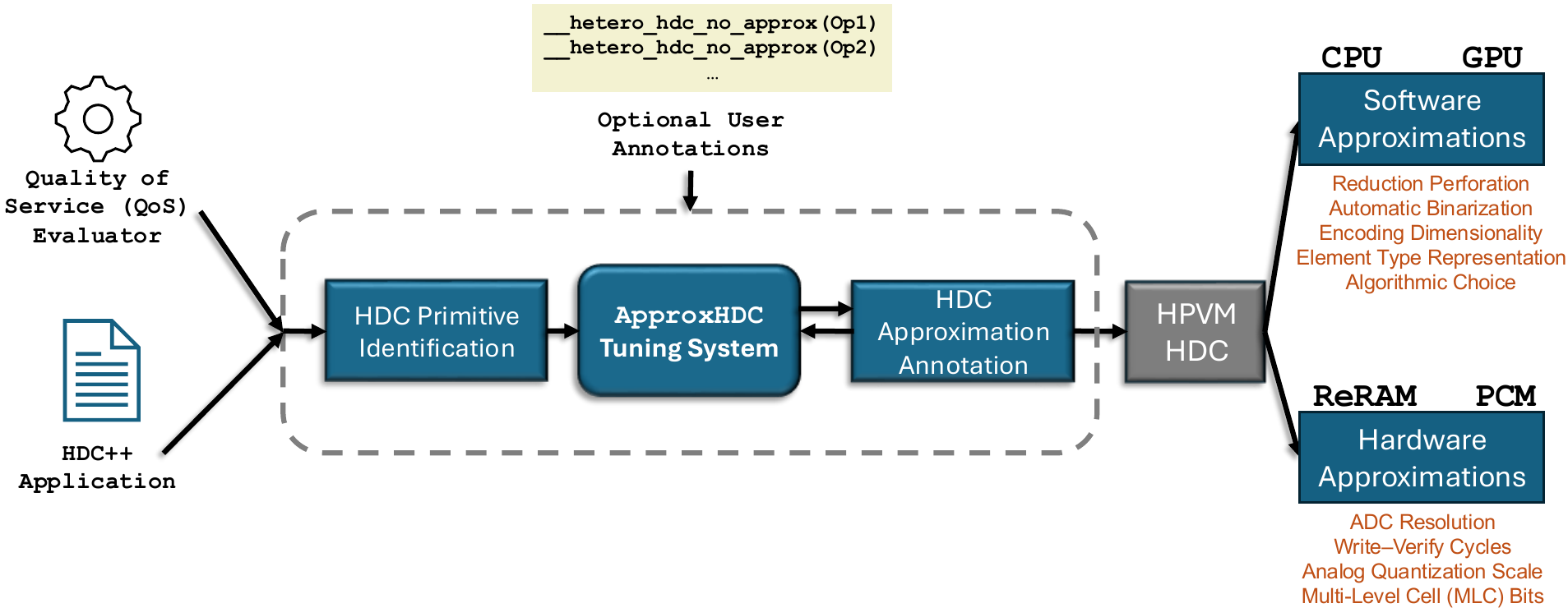}
    \Description{Diagram of the ApproxHDC workflow: HDC++ application and QoS evaluator feed into HDC Primitive Identification, then the ApproxHDC Tuning System explores software and hardware approximation configurations, and finally the HDC Approximation Application pass inserts source-level annotations before compilation and execution.}
    \caption{Overview of the \pname{} workflow. Given HDC applications written in HDC++ along with an application-specific quality-of-service (QoS) evaluator, \pname{} first extracts the set of HDC primitives used in the program. It then performs efficient autotuning to explore approximation choices at both the global and per-operation levels. Finally, \pname{} applies the selected approximations via automatically generated source-level annotations, ensuring that the specified QoS constraints are satisfied. The search space considered by \pname{} spans both software- and hardware-level approximations.}
    \label{fig:overall-workflow}
\end{figure*}
\label{sec:design}
\pname{} is implemented as an extension to the HPVM-HDC compiler through two dedicated compiler passes. Leveraging the approximation specification mechanism provided by HDC++, where approximation types and their parameters can be expressed via in-program source annotations (Section~\ref{bg:hdc++}), \pname{} automatically selects and applies appropriate approximations. Specifically, \pname{} encodes its decisions by inserting the corresponding annotations directly into the HDC++ application, enabling seamless integration with the existing compilation and optimization pipeline. This enables \pname{} to exist externally to the compiler and be easily modified without needing to recompile the HPVM-HDC compiler (which is written in C++). The overall compilation flow is shown in Figure~\ref{fig:overall-workflow}.

\textbf{HDC Primitive Identification.}
The first compiler pass performs a static analysis of the HDC++ application code (prior to the introduction of any approximations) to identify all invocations of HDC primitives used in the program. Each primitive instance is uniquely characterized by its context, which includes the enclosing function and basic block. When multiple occurrences of the same primitive appear within a single basic block, they are further disambiguated based on their execution order within that block. The identified primitives, along with their contextual metadata, are aggregated into a serializable JSON representation, enabling external tools such as \pname{} to access a structured view of the program’s HDC operations (see Listing~\ref{lst:approx_interface}).

\begin{lstlisting}[caption={Interface for \pname{} emitted from HPVM-HDC for approximation tuning.}, label={lst:approx_interface}]
"op": "__hetero_hdc_cos_sim", "Inst_ID": 1, "BB_ID": 8,
    "CallingFunction": "HDCInferenceFunc",
    "ArgType": ["<8 x 1024 x f32>", "<1024 x f32>"],
\end{lstlisting}

\textbf{ApproxHDC Tuning System.}
After identifying the HDC primitives within the application, the \pname{} Tuning System constructs the approximation design space to be explored. This space encompasses both software-level approximations, which are applicable across general-purpose hardware targets such as CPUs and GPUs, and hardware-specific approximations, which target specialized architectures including HD-PCM and HD-ReRAM. The specific types of approximations supported are detailed in Section~\ref{sec:approx-tuning-system}.
To evaluate the impact of different approximation choices, the user provides an application-specific driver that executes the program with the appropriate command-line arguments and datasets, and reports a quality-of-service (QoS) metric (e.g., inference accuracy). The \pname{} Tuning System then empirically explores multiple approximation configurations, measuring QoS, execution time, and energy consumption. \pname{} can be tuned for either end-to-end performance or energy efficiency, subject to a specified QoS constraint.
For each evaluated configuration, the system generates a JSON representation (consistent with the format produced by the HDC Primitive Identification pass) that records the selected approximations and their parameterizations (see Listing~\ref{lst:approx_interface_from_tuner}). Each operation may be assigned zero, one, or multiple approximations. The tuning system further supports user-defined approximation parameters, enabling extensibility to new approximation techniques and hardware targets. By default, the tuning system constructs the approximation space across all HDC operations present in the application. Section~\ref{sec:approx-tuning-system} describes the mechanisms provided by \pname{} to customize and constrain this space, allowing fine-grained control over which operations and approximation parameters are explored.

\begin{lstlisting}[caption={Interface emitted from the \pname{} Tuning System to specify fine-grained, per-operation approximation configurations.}, label={lst:approx_interface_from_tuner}]
"parent_func": "Z4nodePu11matrix_typeIL", 
"intrinsic": "llvm.hpvm.hdc.cossim", 
"number": 0, "op_type": "hmhv", 
"approximations": 
[{"red_perf":  {"base": 1.0, "stride": 8, "extent": 0.84}}, 
 {"alg_choice": "llvm.hpvm.hdc.cossim"}]
\end{lstlisting}
%\vspace{-0.40cm}

\textbf{HDC Approximation Application.}
The second compiler pass consumes the fine-grained approximation configurations produced by the tuning system (as illustrated in Listing~\ref{lst:approx_interface_from_tuner}) and automatically inserts the corresponding source-level approximation annotations into the HDC++ application. Leveraging the contextual metadata collected during the primitive identification phase, each instance of an HDC primitive can be uniquely identified and independently annotated, enabling the application of distinct approximation strategies to different occurrences of the same primitive.

\textbf{Compilation and Execution.}
The HPVM-HDC compiler is employed to perform the necessary compiler transformations specified by the inserted source-level annotations. Section~\ref{sec:approx-tuning-system} describes the set of supported software and hardware approximations, including those developed as part of \pname{}. Using the user-provided application driver, the annotated HDC++ program is compiled and executed, and the corresponding quality-of-service (QoS) metric is recorded. These results are then fed back to the \pname{} Tuning System, which uses them to guide the selection of subsequent approximation configurations. The tuning process continues iteratively until a specified time budget or iteration limit is reached, at which point it outputs the final approximation configuration that maximizes the chosen objective function.
%\vspace{-0.75cm}
\section{Approximation Tuning System}
\label{sec:approx-tuning-system}

\subsection{ApproxHDC Implementation}
\label{sec:approxhdc-impl}
We implement \pname{} in Python for ease of development and integration, in contrast to the HPVM-HDC compiler, which is implemented in C++. \pname{} maintains built-in knowledge of the available HDC primitives and their corresponding approximation transformations in HPVM-HDC. This design choice is essential, as not all approximations are applicable to every HDC operation. For instance, the reduction perforation approximation applies only to reduction-style computations such as cosine similarity.
During operation, \pname{} parses the list of HDC primitives extracted from an HDC++ application (as shown in Listing~\ref{lst:approx_interface}) and determines the set of applicable approximations for each primitive instance. %\xavier{we only use opentuner, does this just raise questions about why we didn't compare to PyATF?: } \ak{Let's remove references to ATF. Opentuner is a pretty popular state-of-the-art autotuner, so people won't question our choice of autotuner.} 
The implementation uses OpenTuner~\cite{opentuner} as the underlying auto-tuning framework. For each tuner, \pname{} automatically constructs the corresponding tuning classes, defining the appropriate parameter spaces and objective functions to guide the search process.

\subsection{Approximation Transformations}
\label{sec:hdc-approximations}
\pname{} provides support for automated approximation tuning of various software and novel hardware approximation for HDC. In this section, we describe the specific approximation transformations which are supported.

\textbf{Reduction Perforation.} Beyond point-wise operations, many HDC primitives perform reduction computations of various kinds. These reductions commonly appear in encoding algorithms such as random projection, which involves matrix–vector dot-product operations, as well as in inference stages, where similarity or dissimilarity metrics are computed through reductions. Owing to the inherent error resilience of HDC algorithms, such computations are amenable to approximation techniques like loop perforation. In reduction perforation, the compiler lowers HDC primitives into loop nests and modifies the reduction-axis loop to skip iterations at regular intervals, defined by a specified stride, rather than iterating over the entire domain. HDC++ exposes this approximation via the \textcolor{blue!41!black}{\texttt{\_\_hetero\_hdc\_red\_perf}} annotation, which accepts three parameters—starting offset, ending offset, and stride—to describe the perforation behavior. For similarity and dissimilarity computations, the resultant values are used as-is; for operations such as matrix–vector dot products, the results are scaled appropriately to compensate for the skipped iterations. 

Due to the diverse ways in which HDC applications are structured, not all reduction operations are equally amenable to perforation. Applying reduction perforation during encoding stages often degrades the quality-of-service (QoS) by producing poorly defined class hypervectors. Moreover, the benefit of perforation is highly application-dependent, as it is influenced by factors such as the encoding dimensionality as well as other approximations within the application. In many cases, performance gains without significant QoS degradation are achieved only beyond an application-specific crossover point, where the reduction size and approximation tolerance align favorably. 

\textbf{Automatic Binarization.} 
% \xavier{this is just an optimization on top of algorithmic choice?}
In the inference stage of HDC, a widely used strategy involves computing the Hamming distance to measure dissimilarity between the input hypervector and a set of class hypervectors. 
% The class hypervector with the lowest dissimilarity is then selected as the predicted label. For hypervectors with integer element types, the Hamming distance is computed by comparing corresponding elements between the input and class vectors, incrementing the distance by one for each mismatch.
A common performance-oriented implementation of Hamming distance in HDC involves preprocessing hypervectors to the $\{+1, -1\}$ representation. This transformation not only improves the QoS for certain applications but also enables use of more hardware-efficient data types. The two possible values can be compactly encoded using packed single-bit representations, with each hypervector element stored as a single bit. This compact format enables highly efficient computation using bitwise operations such as \texttt{xor} and \texttt{popcount}.
% while the final distance values are accumulated using wider integer types.
Moreover, this representation significantly reduces data movement between heterogeneous compute devices such as CPU and GPU. HPVM-HDC automates this inter-procedural transformation using a compiler flag, rewriting both the computation and data movement to leverage the packed representation.

\textbf{Encoding Dimensionality.} 
A key hyperparameter in HDC is the encoding dimensionality, which determines the size of the feature vectors forming the core representation of the algorithm. Increasing this dimensionality can improve the QoS by enabling the model to capture more diverse and discriminative features. However, higher dimensionality also significantly increases computational cost—often quadratically in encoding algorithms and linearly in inference algorithms. This growth not only amplifies overall computation but also intensifies data movement across heterogeneous hardware (e.g., between CPUs and GPUs), leading to additional performance and energy-efficiency degradations.

\textbf{Element Type Representation.}
% \xavier{Need to rework, we only support int32 and fp32, and also binarization. Something about quantization maybe?}
A parameter closely related to the encoding dimensionality is the element type representation used for hypervectors and their associated computations. Hypervectors may be represented using integer data types (8, 16, 32, or 64-bit) or floating-point types (FP16 or FP32), with computations mapped to the corresponding low-level hardware instructions for each data type. In general, HDC training requires higher-precision representations such as floating-point types to ensure stability and accuracy during learning. In contrast, inference computations can often be safely approximated by projecting class and feature hypervectors onto lower-precision integer types. At the extreme, 
%automatic binarization 
We can represent hypervectors using 1-bit integer types, enabling highly efficient bitwise operations while significantly reducing computational and memory overhead.

\textbf{Algorithmic Choice.} 
An HDC application typically comprises computations that can be categorized into three main stages: encoding, training, and inference. Each stage can be realized using multiple algorithmic variants, each offering different trade-offs between performance and QoS. To capture this variability, we incorporate algorithmic choice as an approximation transformation within \pname{}. For instance, HDC inference can be implemented using the computationally expensive cosine similarity metric or the more hardware-efficient hamming distance computation. Similarly, as an example, encoding can be performed using random projection~\cite{rahimi2016robust} or ID-based encoding~\cite{imani2017voicehd}. While \pname{} provides a set of built-in algorithmic choices for common operations, developers can easily extend the framework by defining their own algorithmic variants to suit specific application needs.

\textbf{In-Memory Accelerator Approximations.} In-memory analog accelerators naturally expose multiple approximation knobs stemming from device non-idealities, limited conversion precision, \& stochastic programming behavior. HDC is highly resilient to such noise: prior work shows that even with substantial device-level errors, analog HDC systems retain near–software-equivalent accuracy while achieving large efficiency gains. This robustness enables aggressive approximation in ReRAM/PCM accelerators. We summarize the hardware knobs \& ApproxHDC’s modeling approach:

%\begin{enumerate}[label=\roman*)]
%\item
\noindent
\underline{\textit{( I ) ADC Resolution}:} Analog operations (e.g., current summed dot products) are digitized via ADCs. Lowering ADC bit precision reduces latency and power but increases quantization noise. For example, a 4-bit ADC replaces 256 levels with 16 levels, improving conversion speed at the cost of larger rounding error. \pname{} models this by quantizing the ideal analog value to the target bit-width and adjusting latency/energy accordingly. Accuracy degrades only when quantization noise dominates the HDC similarity signal.\\
    %\item 
\underline{\textit{( II ) Write–Verify Cycles}:} ReRAM and PCM writes rely on iterative program-and-verify steps to reach the target resistance. Fewer verify cycles reduce write latency and energy but increase the chance of under or over-programming, introducing variability in stored values. Our simulator models this by adding resistance noise proportional to the skipped verify steps. These write inaccuracies propagate to analog operations, but HDC's binary and majority computations tolerate noise. Thus, reducing write-verify cycles serves as a practical approximation knob that trades small accuracy loss for significant gains in performance and energy.\\
    %\item 
\underline{\textit{( III ) Analog Quantization Scale}:} Another approximation knob is the quantization scale, which determines how analog outputs are mapped to digital values. By adjusting the sensing range and gain, designers can use a tighter scale that clips or compresses large analog sums for faster, lower-precision sensing, or a wider scale that preserves detail but requires higher-precision ADCs or slower reads. Our simulator models these effects by scaling and capping analog outputs to mimic saturation or reduced sensitivity. The quantization scale thus becomes an accuracy–efficiency lever, and \pname{} varies this parameter to evaluate its impact on sensing energy, performance, and overall HDC accuracy.\\
    %\item 
\underline{\textit{(IV) Multi-Level Cell (MLC) Bits}:} ReRAM and PCM support multi-level cells (MLC) that encode multiple resistance states and thus store several bits per device, increasing density and parallelism. However, higher MLC levels sharply degrade reliability: distinguishing many resistance states introduces significant analog noise and drift, with prior PCM measurements showing more than 10\% error at 3 bits per cell even after write-verify, while 1–2 bits per cell remain far more stable~\cite{SpecPCM2024}. As a result, MLC becomes an approximation knob that trades accuracy for efficiency. Our model reflects this by scaling error probability with bits per cell and incorporating the associated performance benefits (for instance, 2 bits per cell can provide roughly 2× throughput). \pname{} then sweeps these configurations to identify Pareto points, where HDC’s intrinsic noise tolerance often makes 2 bits per cell an attractive balance of accuracy and speed.
%\end{enumerate}

These approximation knobs are exposed through a unified interface in both the compiler and simulator. As in SpecPCM, where the ISA sets bits-per-cell, ADC precision, and verify cycles, our HPVM-HDC backend provides intrinsic hooks for \pname{} to specify these parameters before kernel launch. The simulator then applies calibrated ReRAM/PCM models to inject the corresponding latency, energy, and error behavior. Treating each knob as a dimension in the search space, \pname{} sweeps ADC resolution, MLC level, and write-verify depth to identify Pareto points using end-to-end accuracy and modeled performance/energy. This enables clear trade-off analysis, such as small accuracy loss for meaningful energy savings when lowering ADC bits, versus steeper accuracy penalties of more aggressive MLC configurations.

% \rafae{@Mahbod could you please explain the PCM/ReRAM simulator parameters}
% ADC/Write verify cycles, quantization scales, mlc bits etc.

%\rafae{@Xavier add a code listing showing the generated simulator API with the approximate knobs specified}

 \textbf{Custom Application Specific Parameters.} 
For generality, \pname{} allows application developers to introduce custom, application specific tuning parameters that may not be covered by the existing approximation techniques. Unlike the automatically inferred approximations, these parameters must be explicitly specified by the user. They can represent either integer-valued parameters or enumeration parameters that select among a predefined set of configuration options. This flexibility enables developers to extend the tuning space to include domain-specific controls relevant to their particular HDC application. For example, HD-Clustering, HD-Classification, and RelHD all expose \texttt{TRAIN\_ITERATIONS} as an integer knob governing the number of retraining epochs. HD-HashTable exposes \texttt{KMERS\_PER\_HV}, which controls how many k-mers are packed into each hash-table bucket; smaller values produce more buckets with sparser hypervector sums, changing both the encoding cost and the query-time dot-product distribution. Micro-HD-Classification exposes \texttt{N\_LEVELS}, the number of discrete levels in Level-ID encoding, which controls the quantization granularity applied to continuous input features before they are mapped to hypervectors. RelHD exposes \texttt{RHO\_SHIFT}, which enables a circular left-shift permutation on the level-2 node hypervectors between graph propagation steps; this breaks symmetry between the two aggregation levels so that the similarity computation can distinguish their contributions.

\vspace{-.2 in}
\subsection{Configurable Approximation Tuning Space}
\label{sec:hdc-no-approx}
By default, the tuning system constructs the approximation space over all HDC operations present in an application. A single application may include dozens of distinct HDC primitives, each contributing to the overall computational behavior. 
% \rafae{we should collect the number of HDC ops across applications}.
However, application developers often possess application-specific insight into which operations can tolerate approximation without significantly degrading the QoS metric. While \pname{}’s design could accommodate heuristics for automatically identifying such operations, we instead provide a lightweight, user-driven mechanism for approximation space pruning via source code annotations. Developers can mark specific HDC primitive operations with the \textcolor{blue!41!black}{\texttt{\_\_hetero\_hdc\_no\_approx}} directive to instruct \pname{} to exclude them from approximation. For convenience, users may also annotate entire HPVM-HDC subgraphs with this directive, thereby excluding all enclosed operations from the approximation search space.
This allows users to focus on specific crucial stages within their HDC++ application without needing to perform intrusive changes to the HPVM-HDC compiler or the \pname{} Tuning System. 
%\xavier{where should this next sentence go:?} 
%In almost all of our applications, we use the \texttt{\_\_hetero\_hdc\_no\_approx} directive to 'protect' the Hypervector encoding stage from approximation. Encoding is typically the most sensitive stage of HDC applications, and typically a global reduction in encoding dimensionality has the same effect as approximating encoding.
Nearly all the applications we evaluate \pname{} on require this annotation, specifically to prevent the application specific encoding stages from being approximated. Encoding is typically the most sensitive stage of HDC applications to approximation. 
% \xavier{add discussion of redperf pruning here also?}

\begin{table*}[t]
  \centering
  \small
  \caption{Summary of HDC++ applications used to evaluate \pname{}. All benchmarks are taken from the HPVM-HDC~\cite{hpvm-hdc} suite. Each application defines a domain-specific quality-of-service (QoS) metric that serves as the optimization constraint during automated approximation tuning.}
  \vspace{-0.1 in}
  \label{tab:HDC-workloads}
  \begin{tabular}{@{}lll@{}}
    \toprule
    \textbf{Application} & \textbf{Workload} & \textbf{QoS Metric} \\
    \midrule
    HD-Classification \cite{HD2FPGA}           & Classification using HDC encoding and similarity & Inference Accuracy            \\
    HD-Clustering \cite{HD2FPGA, hdclustering}  & K-means clustering using HDC                     & Normalized Mutual Information  \\
    RelHD \cite{relhd}                          & GNN learning and data relationship analysis           &   Inference Accuracy\\
    HD-HashTable \cite{biohd}                   & Genome sequence search for long reads
sequencing  using HD hashing           &   Inference Accuracy              \\
    \bottomrule
  \end{tabular}
\end{table*}
\section{Evaluation}
\label{sec:eval}

\subsection{Methodology}
\label{sec:eval-methodology}
%\xavier{baselines:, 2-3 sentence summary + baseline sources}

\textbf{Hardware targets.}
The evaluation setup for CPU consists of two AMD EPYC 7453 28-Core processors, with 56 threads and 503GB of memory. 
% \xavier{do we use both sockets??} \ak{Most likely we use one? Not sure.}.
The evaluation setup for GPU consists of an NVIDIA GeForce RTX 2080 Ti GPU with 11GB of memory, on a host cpu of an Intel(R) Xeon(R) Silver 4216 CPU with 16 cores, 32 threads, and 192GB of memory.
% \xavier{nusrat CPU, miranda GPU.}

\textbf{Applications.}
%We implement four applications in HDC++. \xavier{how to anonymously describe baselines / porting? } 
%\xavier{Where to mention MicroHD version of HD-Classification?} \ak{If it is open source, provide a link in the footnotes. If not, then just say that we contacted the authors and they provided the code to us, but it is not publically available.} 
% \xavier{I think we might want to merge Hd-Classification and HPVM version of Micro HD's application}
We evaluate \pname{} using benchmarks from the HPVM-HDC~\cite{hpvm-hdc} suite, which provides baseline implementations in HDC++ without any approximation annotations. Table~\ref{tab:HDC-workloads} summarizes the application workloads and the corresponding datasets used in our evaluation.
%We evaluate \pname{}'s automated approximation tuning system using four applications from HPVM-HDC\cite{hpvm-hdc} benchmark suite. 
\textbf{HD-Classification} implements a classification algorithm using HD encoding and similarity primitives. The QoS (quality of service) metric for HD-Classification is classification accuracy. \textbf{HD-Clustering} implements K-Means clustering using HDC. HPVM-HDC exposes two independently tunable similarity primitives, one for training and one for inference. The QoS metric we use is NMI (Normalized Mutual Information) score. We evaluate HD-Clustering and HD-Classification on the ISOLET dataset of spoken letter recordings \cite{isolet}.  \textbf{RelHD} \cite{relhd} implements a GNN learning algorithm using HDC. It encodes node and edge features as hypervectors, and uses similarity metrics to produce node classifications. We evaluate on the Cora dataset of categorized scientific publications \cite{cora}. \textbf{HD-HashTable} \cite{biohd} implements a genomic sequence search using HDC. Reference k-mers from a genome are bundled in multi-HV buckets. At query time, the nearest hypervector is determined via dot-product similarity. The similarity threshold and k-mers per Hypvervector are both exposed as tunable knobs.

%\vspace{-.2 in}
\subsection{Tuning Space Pruning} 
\label{sec:eval-pruning}
To demonstrate the effectiveness of the HDC-specific techniques to prune the approximation search space, we run and evaluate the \pname{} system across all four HDC++ applications, on both CPU and GPU. We use four different search space configurations, each demonstrating a more aggressive form of pruning. For all configurations all other application knobs are enabled unless otherwise mentioned. The different search space configurations are as follows:

%\begin{enumerate}
    %\item 
\vspace{0.1 in}
\noindent
\underline{\textbf{(1)} \textbf{Full RedPerf}:} all three reduction perforation knobs are exposed---start, stride, and end---alongside encoding and data-type knobs. This is the most expressive condition but introduces the largest search space.

\noindent
\underline{\textbf{(2)} \textbf{End-only + no approx hint enabled}:} only the end (extent) knob is tunable; start and stride are fixed at their baseline values. This substantially reduces the search space while retaining the most impactful approximation lever (truncating the reduction). 
% \xavier{\cite{DPQ-HD}}

\noindent
\underline{\textbf{(3)} \textbf{No RedPerf}:} reduction perforation is disabled entirely. The tuner varies only global knobs (encoding dimension) and application-level parameters such as training iterations.

\noindent
\underline{\textbf{(4)} \textbf{Full RedPerf + no approx hint enabled}:} all three redperf knobs are tunable as in condition~(1), but the search space is additionally pruned by the no approx hint.
%\end{enumerate}
\vspace{0.1 in}

\begin{figure*}[t!]
    \centering
    \includegraphics[width=\linewidth]{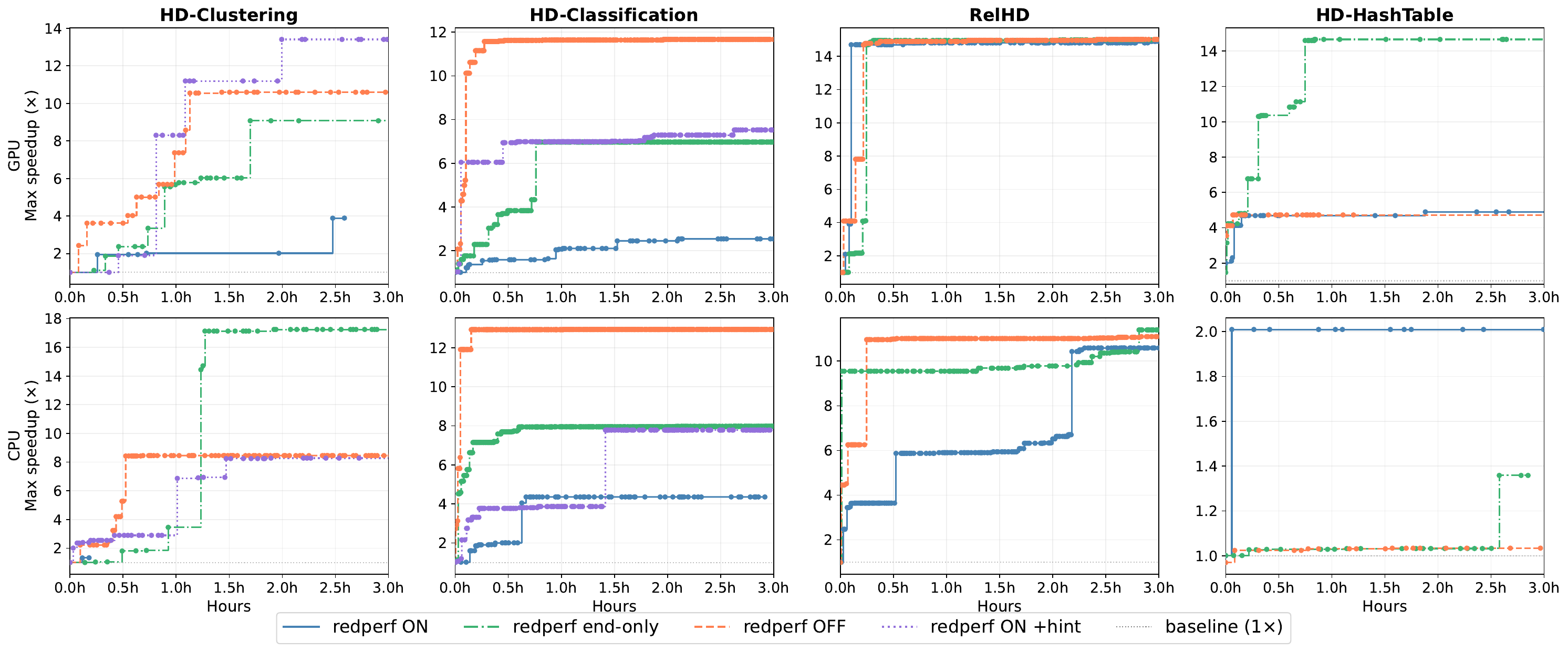}
    \Description{Matrix of line plots showing running-best speedup over tuning time for all applications under four reduction perforation pruning conditions, split by GPU (top row) and CPU (bottom row).}
    \vspace{-0.3in}
    \caption{Maximum speedup achieved over the 3-hour tuning budget for all four
             reduction perforation pruning conditions across all applications and GPU top row, CPU bottom row. Each curve shows the
             running-best speedup relative to the untuned baseline as new
             configurations are evaluated. End-only with encoding protection
             (green) typically finds more valid configurations and reaches
             higher peak speedups than full reduction perforation (blue), whose
             larger search space impedes exploration within the fixed budget.
             }
    \vspace{-1em}
    \label{fig:redperf-matrix}
\end{figure*}

We evaluate the pruning effectiveness by running ApproxHDC for each configuration using a 3 hour tuning budget. The baseline configuration is the unapproximated version of the application.  The target QoS metric used for each tuning run is the baseline QoS minus 0.05. All QoS metrics are measured from 0 to 1.0. This represents a threshold of $5\%$ accuracy or 0.05 difference in NMI score. We seed each auto-tuning run with the baseline configuration as a start point. Figure ~\ref{fig:redperf-matrix} shows the running-best speedup achieved over the 3-hour tuning budget for each pruning condition, across all four applications on GPU (top row) and CPU (bottom row). Shown configurations are subject to QoS constraints.

% \paragraph{Tuning Configurations.} 
% \xavier{Describe what tuning configurations are best and why (end-only) usually wins.}
% The end-only condition consistently achieves the highest or near-highest speedup across all applications while also finding the largest number of valid configurations within the budget. On HD-Clustering (GPU), end-only reaches a peak speedup of 22.4$\times$ versus 11.4$\times$ for full redperf. On RelHD (GPU), end-only achieves 28.5$\times$ speedup  compared to 14.5$\times$ for full redperf. On HD-HashTable, end-only reaches 4.4$\times$ speedup. HD-Classification shows more modest gains across all conditions (1.4--1.8$\times$). \xavier{as only two HPVM-level similarity operations are exposed to the tuner.}

% The advantage of end-only over full redperf is attributable to search space size. Table~\ref{tab:placeholder}\rafae{broken ref} shows that exposing all three redperf knobs inflates the search space by several orders of magnitude (e.g., $10^{69.72}$ vs.\ $10^{22.58}$ for HD-Clustering). Within a fixed budget, OpenTuner explores a smaller fraction of the full-redperf space, resulting in fewer valid configurations found. End-only retains the most impactful knob---the extent of the reduction---while eliminating the start and stride dimensions that add search complexity. 

% \textbf{Best Configurations / Knob Evaluation.} 

\textbf{HD-Clustering.}
The baseline configuration uses float data type, cosine similarity, 20 Training iterations, and a dimensionality of 1024, achieving a Normalized Mutual Information (NMI) score of 0.72. \pname{} was able to find 13.41x (0.6708 NMI score) speedup on GPU, and 17.25$\times$ (0.6737 NMI score) speedup on CPU. On GPU, this speedup comes from switching to int data type and using hamming similarity, and reducing training iterations down to 3. The reduction perforation extent was held at 1.0=full. On CPU, this speedup comes from the same configuration changes, but also reducing reduction perforation inference extent to 0.9. Both devices held D=1024. For both devices, the redperf ON series struggles to find high speedups. HD-Clustering tuning runs were configured with a target NMI score of .62, below the .67 NMI score threshold that represents a reasonable QoS tradeoff from the baseline QoS score. This causes a large portion of the search budget to be wasted on evaluating configs below the QoS cutoff. The results for HD-Clustering represent a conservative estimate of max speedup within the 3 hour search budget. %\xavier{For HD-Clustering, the tuning run was done with a target accuracy of 0.62, while the plot and evaluation use a cutoff of 0.67 NMI, so the difference between series doesn't really make sense to evaluate. Do we need to explain this. I think we only need to explain it if we need to explain the tuning series graphs.}. 

\textbf{RelHD.}
For RelHD, we only evaluate three tuning-series, as the encoding method used is not approximate, making the no-approx hint a no-op. The baseline configuration uses D=2048, cosine similarity, and float data type. RelHD includes custom 'Rho-Shift knob', which is a RelHD encoding specific algorithmic choice, representing a circular shift of hypervectors in a portion of the encoding stage. The main speedup on both CPU and GPU comes from reducing hypervector dimensionality from 2048 to 128 (16x reduction). On CPU, we were able to achieve an 11.4x speedup (.71 accuracy), and on GPU we were able to achieve a 15.02x speedup (.71 accuracy). Among top-performing configurations similarity ops saw little approximation by reduction perforation settings, and cosine similarity / float was chosen. Approximating RHO\_SHIFT had little effect on performance and accuracy. All three tuning series, for both CPU and GPU, were able to converge on the same configuration area and speedup. On CPU, the series that enable reduction perforation are able to achieve a slightly higher (.3x) speedup than the redperf OFF series. On GPU, the reduction perforation does not have a significant effect on the performance or accuracy, seen by all series plateauing at the same performance, so the tuner does not spend much of its budget searching across redperf knobs, unlike CPU.

\textbf{HD-Classification.}
For HD-Classification, the baseline configurations is D=2048, Train Iterations = 10, using float datatype and cosine similarity, achieving a .87 accuracy. Both CPU and GPU reach the same best performing configuration, by reducing dimensionality down to 256 (8x), and training iterations down to 3. CPU achieves a speedup of 11.64x, and GPU achieves a speedup of 12.93x, both get (.82 accuracy). Both maximum speedup configurations were found by the redperf OFF series. Enabling reduction perforation for HD-Classification increases the search space size, and prevents the tuning run from finding the high performing configurations within a 3-hour time budget. 
% \xavier{this is just a completely different story from Micro-HD-Classification, which is confusing.}

\textbf{HD-HashTable.}
For HD-HashTable, the baseline configuration is $D=8192$, $\text{KMERS\_PER\_HV}=400$, \\ $\text{THRESHOLD\_MULT}{=}0.9$, using float data type and dot-product similarity. The baseline achieves an inference accuracy of .83. On GPU, the end-only series achieves $14.66\times$ speedup (0.78 accuracy) by reducing $D$ from 8192 to 512, setting, $\text{KMERS\_PER\_HV}{=}20$ (sparser hash-table buckets), \\ $\text{THRESHOLD\_MULT}{=} 0.589$, and the redperf extent to $0.646$ (using only 65\% of the hypervectors for the query dot product). The Full RedPerf and No RedPerf series both plateau at around $5\times$, neither tune the dimensionality enough to gain similar speedups. On CPU, the speedups are much lower, getting only up to  $2.01\times$ (Full RedPerf). 

The search space for HD-Hashtable is substantially larger than the other applications at the No RedPerf pruning series, ($10^{16}$ vs. $10^{3}$--$10^{4}$ for the other apps). The runtime of HD-HashTable is also much longer than the other applications, especially on CPU. These factors combined means on CPU the search process is severely limited by the tuning budget, only evaluating $< 100$ configurations within the 3 hour limit. 
% On CPU, the Full RedPerf series reaches $2.01\times$ while End-only and No RedPerf are stuck near $1\times$; all three series sampled $D{=}4096$ configurations, but only Full RedPerf's attempt kept  and passed the accuracy bar. 

\subsection{Comparison to MicroHD}
MicroHD~\cite{microhd-github} implements its own version of classification using HDC in Python with limited approximation tuning support. To compare tuning systems, we port this version to HDC++ as Micro-HD-Classification. Micro-HD uses Level-ID encoding, which exposes a Level integer knob controlling the number of levels in Level-ID encoding (range 2--1024) and a Quantization knob controlling the hypervector bit width (1--16 bits). %\rafae{this sentence is redundant; just say we support it as a custom paramater}ApproxHDC exposes $L$ through its custom-application-parameter mechanism: the application registers $L$ as an \texttt{ApproxCustomInteger} knob. 
\pname{} supports tuning for this through a custom tuning parameter. HDC++ does not support sub-word hypervector quantization, so our HDC++ port of Micro-HD-Classification does not expose $Q$; we allow MicroHD to tune $Q$ when measuring MicroHD's speedup.

To compare the two tuning systems we fix a shared reference configuration of $D{=}1024$, $L{=}128$, $Q{=}16$, using cosine similarity. Each system is evaluated in its own native execution model: \pname{} speedups compare an HPVM-compiled tuned binary against the HPVM-compiled reference, and MicroHD speedups compare its best PyTorch configuration against the PyTorch reference. The two backends report nearly identical baseline accuracy on ISOLET ($0.8954$ on HPVM, $0.8972$ on PyTorch).
 We measure the maximum speedup (under accuracy threshold constraint), achieved by both \pname{} and MicroHD on GPU.

%\begin{wraptable}{r}{10cm}
% \begin{table}[t]
% \centering
% \small
% \caption{speedups. \xavier{table width?}\rafae{Can this be inlined into the text itself. We can save a lot of space then}}
% \vspace{-0.1 in}
% %\resizebox{2 in}{!}{%
% \footnotesize
% \begin{tabular}{@{}lcc@{}}
% \toprule
% System            & Speedup                & Accuracy \\
% \midrule
% MicroHD (PyTorch) & $7.07\times$           & $0.871$ \\
% ApproxHDC (HPVM)  & $\mathbf{24.65\times}$ & $0.847$ \\
% \bottomrule
% \label{table:microhd}
% \vspace{-0.75cm}
% \end{tabular}
% %}%
% \end{table}
%\end{wraptable}

\begin{figure}[t]
    \centering
    \includegraphics[width=\linewidth]{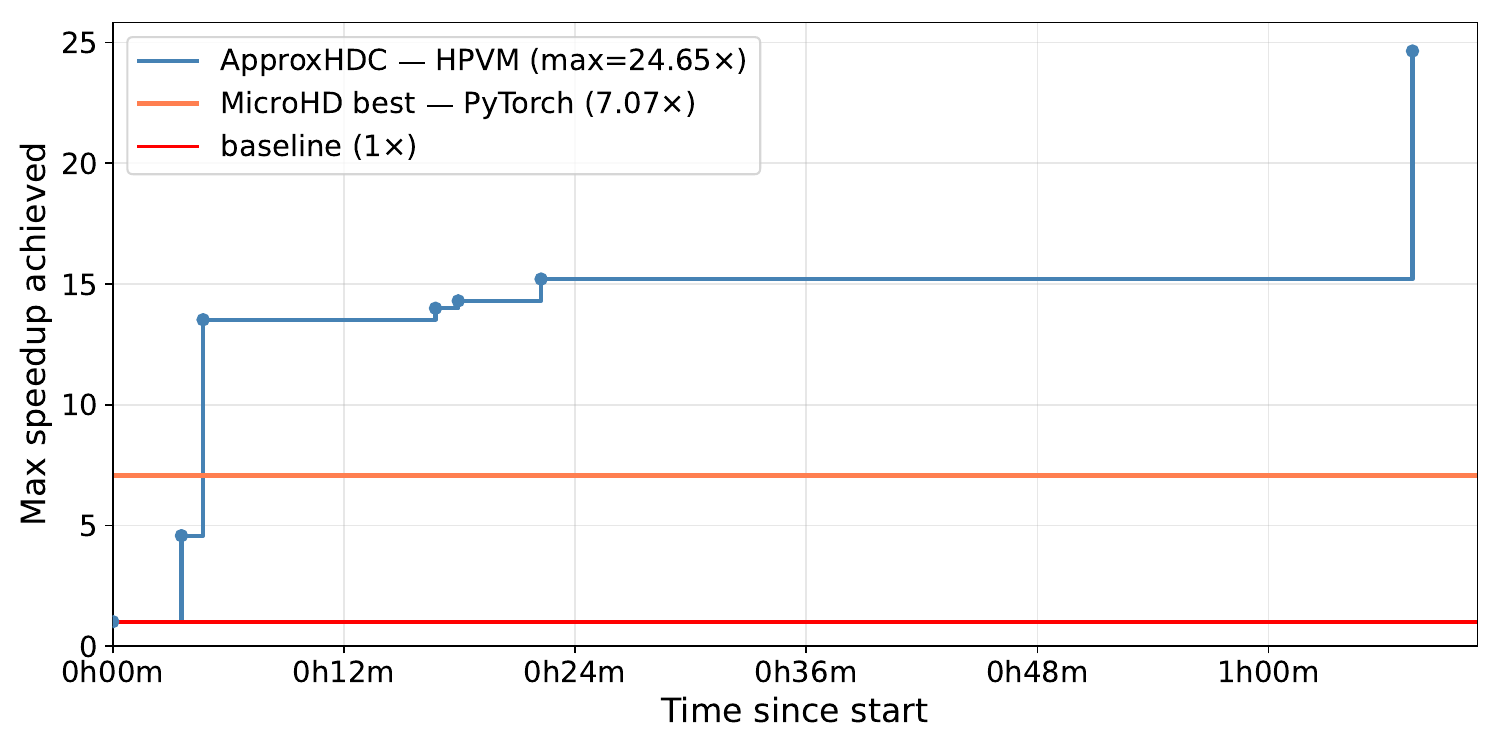}
    \Description{Line plot comparing ApproxHDC running-best speedup over tuning time against MicroHD's fixed speedup baseline for the classification benchmark, showing ApproxHDC surpassing MicroHD within approximately 5 minutes.}
    \vspace{-0.3in}
    \caption{
    %\xavier{Beats MicroHD at T=4m41s, MicroHD takes 4m13s} 
    Measured maximum speedup over time for MicroHD’s classification benchmark using \pname{}’s approximation tuning framework. MicroHD’s achieved speedup is shown as a horizontal reference line. \pname{} identifies superior approximation configurations within 5 minutes and ultimately discovers a configuration that is approximately 3.5× more efficient than MicroHD. } 
    \label{fig:microhd-speedup}
    \vspace{-1.5em}
\end{figure}

\pname{} additionally exposes Reduction Perforation knobs (\textit{base}, \textit{stride}, \textit{extent}) on every HDC operator; we fix these to their no-perforation values ($\text{extent}{=}1.0$, $\text{stride}{=}1$, $\text{base}{=}0$) in the reference config. \pname{} also tunes the encoder algorithmic-choice knob, permitting Random Projection in place of Level-ID. MicroHD is run until its binary search saturates ($\sim$300s on ISOLET); \pname{} is given a fixed 2-hour tuning budget. Both systems use a shared accuracy cutoff and target of $0.845$. Results are shown in Figure~\ref{fig:microhd-speedup}:

\pname{} finds an approximation configuration that results in $24.65\times$ speedup, with a 4.8\% loss in accuracy. MicroHD finds configurations with up to a $7.07\times$ speedup and a 2.4\% loss in accuracy. \pname{} outperforms MicroHD's best accepted configuration by $3.49\times$ at the same accuracy bar. The systems converge on structurally different solutions:

%\begin{itemize}
\pname{} picks $D{=}512$ (a $2\times$ reduction) but aggressively perforates both cossim operators, sampling only $\sim$44\% of the reduction length during training and $\sim$82\% at inference. It sets $L{=}814$ \& selects Level-ID encoding with cosine sim. 

% \xavier{explain why high L doesn't matter?}

MicroHD picks $D{=}1000$ (no meaningful $D$ reduction from the $1024$ baseline) and instead cuts levels, from $128$ down to $32$ ($4\times$), while leaving quantization at the full $Q{=}16$. MicroHD has no analogue of per-op reduction perforation, so it cannot chase the extent knobs that drive ApproxHDC's speedup.
%\end{itemize}

The $3.49\times$ gap is a result of ApproxHDC's more expressive search space: the tuner finds that RP-extent on both cossim operators is load-bearing at strict accuracy, and those axes do not exist for MicroHD. \pname{} crosses MicroHD's max speedup very early in its budget: the first \pname{} configuration to beat MicroHD's best lands at $t{\approx}282$~s ($\sim$4.7~min) into tuning, already at $13.52\times$. The remaining $\sim$2h of the \pname{} budget is spent increasing $13\times$ up to $24.65\times$.

\subsection{SpecPCM}
Beyond software approximation on CPUs and GPUs, \pname{} supports hardware-level approximation knobs exposed by the SpecPCM in-memory computing accelerator (Section~\ref{sec:approx-tuning-system}). We evaluate \pname{} on SpecPCM for HD-Classification using the SpecPCM simulator~\cite{SpecPCM2024}. Tunable hardware knobs include ADC resolution, write-verify cycles, quantization scale, and multi-level cell (MLC) bits per device.

\begin{figure}[t]
    \centering
    \includegraphics[width=\linewidth]{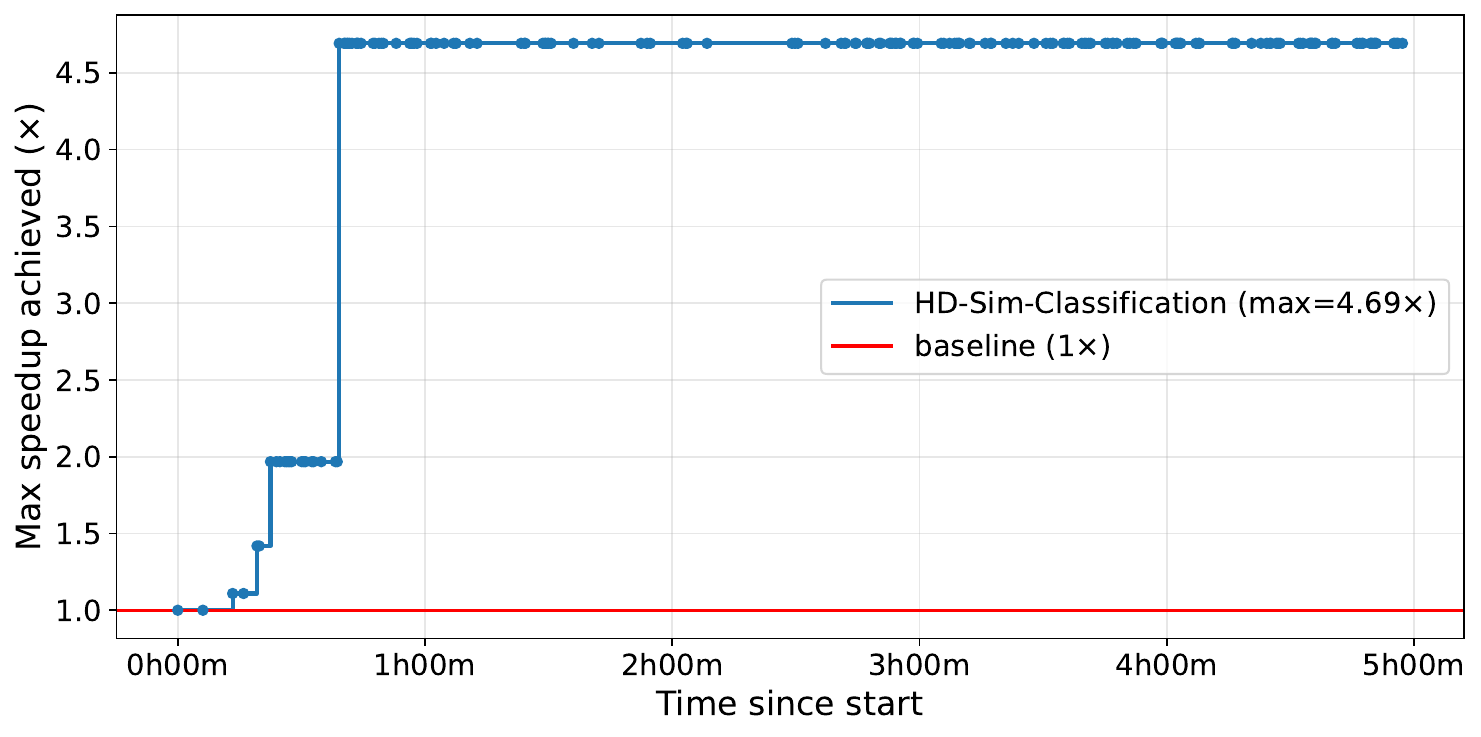}
    \Description{Line plot of running-best speedup over tuning time for ApproxHDC hardware approximation on the SpecPCM Phase Change Memory simulator, reaching over 4.5x speedup within 45 minutes.}
    \vspace{-0.3 in}
    \caption{ \pname{} hardware approximation tuning of the HD-Classification benchmark for the SpecPCM~\cite{SpecPCM2024} Phase Change Memory simulator. \pname{} effectively achieves over 4.5x performance improvements within 45 minutes of tuning (and simulation) time.}
    \label{fig:sim-speedup}
    %\vspace{-1.10cm}
\end{figure}

We use a reference configuration of $D{=}1024$, \\$\text{TRAINING\_ITER}{=}10$, $\text{MLC}{=}1$, $\text{ADC\_RES}{=}8$, $\text{ADC\_STEP}{=}1$, $\text{ADC\_FLOOR}{=}{-}127$, $\text{ADC\_CEIL}{=}128$, \\$\text{WRITE\_VERIFY\_CYCLES}{=}0$. On the simulator this configuration achieves an accuracy of 0.84. We run \pname{} for a 3-hour tuning budget against an accuracy cutoff and target of .79. The complete tuning time trace is shown in Figure ~\ref{fig:sim-speedup}.

The best configuration sets $D{=}1024$, $\text{TRAINING\_ITER}{=}2$, $\text{MLC}{=}1$, $\text{ADC\_RES}{=}8$, $\text{ADC\_STEP}{=}2$, $\text{ADC\_FLOOR}{=}{-}3$,\\  $\text{ADC\_CEIL}{=}128$, $\text{WRITE\_VERIFY\_CYCLES}{=}0$. This configuration achieves an accuracy of .8138. The majority of the 4.69$\times$ speedup is driven by cutting training iterations from 10 to 2. While the SpecPCM hardware knobs (MLC, ADC resolution, ADC step, ADC floor, ADC ceiling) do not have a significant effect on performance, they do impact device energy usage. We did not tune for energy explicitly, the tuning objective was set to maximize runtime speedup, the tuner's best-speedup configuration already achieves a $4.69\times$ energy reduction alongside its $4.69\times$ speedup, \& we observed SpecPCM configurations that trade a small additional accuracy loss of .55\% for significantly higher energy reduction (up to $7.12\times$) by dropping $\text{ADC\_RES}$ from 8 to 6 \& widening $\text{ADC\_STEP}$ to 9. Notably, the tuner converges on a strongly asymmetric ADC window: $\text{ADC\_FLOOR}$ rises from the baseline ${-}127$ to $\sim{-}3$ across all top configurations, while $\text{ADC\_CEIL}$ stays at 128.  The SpecPCM simulator computes dot-product similarity over 128-element array chunks, with each partial dot product quantized by the ADC before accumulation. At $\text{MLC}{=}1$ with $\pm 1$ encoded hypervectors, the partial sums for the correct class are predominantly positive (similar vectors yield positive inner products), while negative partial sums appear mainly for non-matching classes. Clipping below ${-}3$ therefore preserves the classification outcome while narrowing the ADC input range. This narrower range allows fewer quantization levels (lower $\text{ADC\_RES}$, wider $\text{ADC\_STEP}$) to cover the active range, driving  energy reduction.

\begin{table}[t]
\centering
\caption{Approximate search-space sizes for each pruning
  configuration. Pruning reduces the space by up to 86 orders of magnitude
  (e.g., HD-Classification: $10^{89} \to 10^{3}$). N/A indicates the
  configuration was not evaluated.}
\vspace{-0.1 in}
\label{tab:search-space}
\resizebox{\columnwidth}{!}{%
\footnotesize
\begin{tabular}{@{}lrrrr@{}}
\toprule
\textbf{Pruning Config.}
  & \textbf{HD-Class.}
  & \textbf{HD-Clust.}
  & \textbf{RelHD}
  & \textbf{HD-Hash.} \\
\midrule
Full RedPerf        & $10^{89}$ & $10^{70}$ & $10^{27}$ & $10^{26}$ \\
End-only + hint     & $10^{41}$ & $10^{23}$ & $10^{24}$ & $10^{25}$ \\
No RedPerf          & $10^{3\phantom{0}}$  & $10^{3\phantom{0}}$  & $10^{4\phantom{0}}$  & $10^{15}$ \\
Full RedPerf + hint & $10^{47}$ & $10^{25}$ & N/A       & N/A       \\
\bottomrule
\end{tabular}%
}
%\vspace{-2em}
\end{table}

\section{Related Work}

\textbf{Compiler IR for Approximations.}
EnerJ~\cite{enerJ} supports approximate computing via type declarations that mark data as precise or approximate, enabling static analysis to isolate precise and approximate code; hardware-level approximations include reduced DRAM refresh rates, lower SRAM voltage, and approximate floating-point operations. While effective for general-purpose applications, EnerJ’s design limits its applicability to domain-specific workloads such as HDC and does not support heterogeneous accelerator targets. Rely~\cite{rely} provides a programming language and analysis framework that enables developers to specify quantitative reliability requirements for programs running on unreliable hardware (i.e., hardware where arithmetic operations or memory accesses may fail with some probability). ApproxHPVM~\cite{approxhpvm} and ApproxTuner~\cite{approxtuner} provide compiler- and runtime-level support for accuracy-aware approximations in deep learning and image processing. ApproxHPVM encodes per-operation error metrics in the HPVM IR and maps computations to approximate hardware to meet end-to-end quality constraints. %, achieving up to 9× speedup and 11× energy reduction. 
ApproxTuner extends this with a three-stage tuning framework that explores approximation-performance trade-offs at development, install, and runtime, achieving significant speedups with minimal accuracy loss. Both frameworks are effective for functional, tensor-based workloads but do not directly support domain-specific paradigms like hyperdimensional computing (HDC), which require more intrusive approximation techniques to maintain acceptable quality-of-service given their inexpensive, non-backpropagation-based learning algorithms. %Moreover, HDC applications rely on specialized HDC operations interleaved with general-purpose parallel C++ code, making domain-specific compiler and approximation support essential.

%ApproxHPVM
%Rely
%TVM
%enerj

\textbf{Hyperdimensional Computing Approximation.}
%MicroHD~\cite{microhd} is an accuracy-driven optimization framework for hyperdimensional computing targeting TinyML and edge environments. It explores a limited hyperparameter-based approximation space, including encoding dimensionality, quantization bit-width, number of hypervectors, and encoding algorithm choice, which is smaller compared to frameworks that support more diverse or fine-grained approximations. MicroHD employs a combination of binary search and greedy algorithms to efficiently select configurations that minimize memory and compute requirements while satisfying user-specified accuracy constraints, achieving up to 200–$266\times$ resource savings with less than 1\% accuracy loss. The framework is implemented using TorchHD~\cite{torchhd}, a Python library for HDC, and does not leverage a compiler IR; as a result, all transformations and approximations must be manually implemented, and it lacks flexibility in defining application-specific tuning parameters or hardware-specific knobs. 
MicroHD~\cite{microhd} is an accuracy-driven HDC optimization framework for TinyML and edge settings that explores a limited hyperparameter-based approximation space (e.g., dimensionality, quantization, and encoding choices). It uses binary search and greedy methods to meet accuracy constraints while reducing resource usage, achieving up to 200–266× savings with <1\% accuracy loss. Built on TorchHD~\cite{torchhd} without a compiler IR, it requires manual implementation of approximations and offers limited flexibility for application- or hardware-specific tuning.
In contrast, our approach is extensible, as it automatically constructs the approximation space, allows the definition of custom application-specific tuning parameters with ease, and supports hardware approximation knobs not previously available, while also providing compiler-based insertion and management of software- and hardware-level approximations across heterogeneous targets.~\cite{kazemi2022achieving} show that HDC can tolerate hardware-level approximation using multi-bit FeFET in-memory computing, achieving software-equivalent accuracy with up to $826\times$ energy and 30× latency improvements. This demonstrates that HDC is resilient to approximate hardware execution, motivating compiler-based frameworks to systematically exploit such hardware knobs for performance and efficiency gains.

%% No need for the following. we are not proposing an autotuner.
%\textbf{Approximation Tuning Frameworks.} OpenTuner~\cite{opentuner} is an extensible autotuning framework that automates exploration of large configuration spaces using an ensemble of search techniques. It allows developers to define custom search spaces, objectives, and search algorithms, including a multi-armed bandit strategy to allocate more evaluations to promising search techniques while reducing time spent on underperforming ones. This approach enables efficient optimization across diverse domains and hardware targets. OpenTuner has been applied to performance tuning, multi-objective optimization, and compiler parameter selection, providing a flexible foundation for frameworks that require systematic exploration of program parameters, including in approximate or domain-specific computing. pyATF~\cite{pyatf} is a Python-based auto-tuning framework that supports constraint-aware parameter spaces, enabling efficient pruning of invalid configurations. It employs multi-armed bandit algorithms to allocate search effort to promising strategies, similar to OpenTuner, but with built-in handling of interdependent parameters. Compared to OpenTuner, pyATF is better suited for domains with complex parameter relationships, such as compiler flags or hardware knobs, while OpenTuner offers more flexible, general-purpose search and custom search heuristics.

\section{Conclusion}
Hyperdimensional computing (HDC) is robust to approximation but hard to tune. We present \pname{}, an automated framework that explores hardware and software approximations while optimizing QoS, with optional user guidance. Results show no single strategy consistently maximizes speedup, highlighting the need for guided tuning.

% Acknowledgements only in accepted papers! 
\begin{comment}
\begin{acks}
   This document is derived from previous conferences, in particular MICRO 2013, ASPLOS 2015, MICRO 2015-2025, ISCA 2025, as well as SIGARCH/TCCA's Recommended Best Practices for the Conference Reviewing Process. 
\end{acks}
\end{comment}

%%%%%%% -- PAPER CONTENT ENDS -- %%%%%%%%

\bibliographystyle{ACM-Reference-Format}
\bibliography{refs}

\end{document}